\documentclass[aps,superscriptaddress,twocolumn]{revtex4-1}
\usepackage{amssymb,amsmath,mathptmx}
\usepackage{graphicx}
\usepackage{dcolumn}
\usepackage{bm}
\usepackage{hyperref}
\usepackage{color}
\usepackage[utf8x]{inputenc}
\usepackage{array}
\newcolumntype{C}{>{\centering\arraybackslash}p{1em}}

\newcommand{\be}{\begin{equation}}
\newcommand{\ee}{\end{equation}}
\newcommand{\bea}{\begin{eqnarray}}
\newcommand{\eea}{\end{eqnarray}}
\begin{document}
\title{A quantum interferometer for quartets in superconducting three-terminal Josephson junctions}

\date{\today}

\author{R\'egis M\'elin}
\affiliation{Universit\'e Grenoble-Alpes, Institut N\'eel, BP
  166, F-38042 Grenoble Cedex 9, France}
\affiliation{CNRS, Institut N\'eel, BP 166, F-38042 Grenoble
  Cedex 9, France}
\author{Denis Feinberg}
\affiliation{Universit\'e Grenoble-Alpes, Institut N\'eel, BP
  166, F-38042 Grenoble Cedex 9, France}
\affiliation{CNRS, Institut N\'eel, BP 166, F-38042 Grenoble
  Cedex 9, France}

\begin{abstract}
  An interferometric device is proposed in order to analyze the quartet
  mode in biased three-terminal Josephson junctions (TTJs), and to
  provide experimental evidence for emergence of a single stationary
  phase, the so-called quartet phase. In such a quartet-Superconducting
  Quantum Interference Device (quartet-SQUID), the flux sensitivity
  exhibits period ${hc}/{4e}$, which is the fingerprint of a transient
  intermediate state involving two entangled Cooper pairs. The
  quartet-SQUID provides two informations: an amplitude that measures a
  total ``quartet critical current'', and a phase lapse coming from the
  superposition of the following two current components: the quartet
  supercurrent that is odd in the quartet phase, and the phase-sensitive
  multiple Andreev reflection (phase-MAR) quasiparticle current, that is
  even in the quartet phase. This makes a TTJ a generically
  "$\theta$-junction". Evidence for phase-MARs plays against
  conservative scenarii involving synchronization of AC Josephson
  currents, based on ``adiabatic'' phase dynamics and RSJ-like models.
\end{abstract}
\maketitle

{\it Introduction: } Multiterminal Josephson junctions (MTJs)
\cite{Omelyanchouk1,Omelyanchouk2,Omelyanchouk3,Omelyanchouk4} appear
as a very fertile evolution in the field of superconductivity. While
unbiased MTJs offer prospects as platforms for controllable
topological properties
\cite{vanHeck,Padurariu,Nazarov1,Nazarov2,topo0,Feinberg1,Feinberg2,topo1,topo2,topo3,topo4,topo4-bis,topo5,Berry,topo1-plus-Floquet,Levchenko1,Levchenko2,Akh2,Gavensky},
biased MTJs reveal new channels for both superconducting
phase-sensitive and quantum mechanical DC currents, as predicted by
theory
\cite{Cuevas-Pothier,Houzet-Samuelsson,Freyn,Melin1,Jonckheere,biSQUID,FWS,Sotto,engineering,Nowak,paperI,paperII,long-distance,PRB2022,Akh2,Pillet,EPJB,Pillet2,Nazarov-PRR,Nazarov-PRB-AM}
and confirmed in experiments
\cite{Lefloch,Heiblum,HGE,multiterminal-exp1,multiterminal-exp2,multiterminal-exp3,multiterminal-exp4,multiterminal-exp5,multiterminal-exp6,multiterminal-exp7,multiterminal-exp8,multiterminal-exp9,multiterminal-exp10}. A
paradigm of multiterminal Josephson junction \cite{Freyn} involves
three superconductors biased at the opposite voltages $0,V,-V$, this
making the junction host Cooper quartets \cite{Freyn}. Those transient
quartets are made of entangled pairs of Cooper pairs and flowing from
the unbiased terminal towards the two others simultaneously. This
voltage configuration ensures energy conservation, a necessary
condition for having DC Josephson currents. The quartet mechanism goes
together with emergence of a stationary phase combination of the three
terminal phases, the so-called quartet phase $\varphi_Q$. At the
microscopic level, the minimal process appearing in perturbation
theory in the tunnel amplitudes consists of four Andreev
reflections. Quartets (as well as higher-order multipairs such as
sextets, octets, ....)  therefore constitute a genuine quantum
mechanical mesoscopic phenomenon, not occurring in simple classical
Josephson arrays but instead in truly multiterminal junctions.

\begin{figure*}[t]
  \begin{minipage}{.6\textwidth}
    \includegraphics[width=\textwidth]{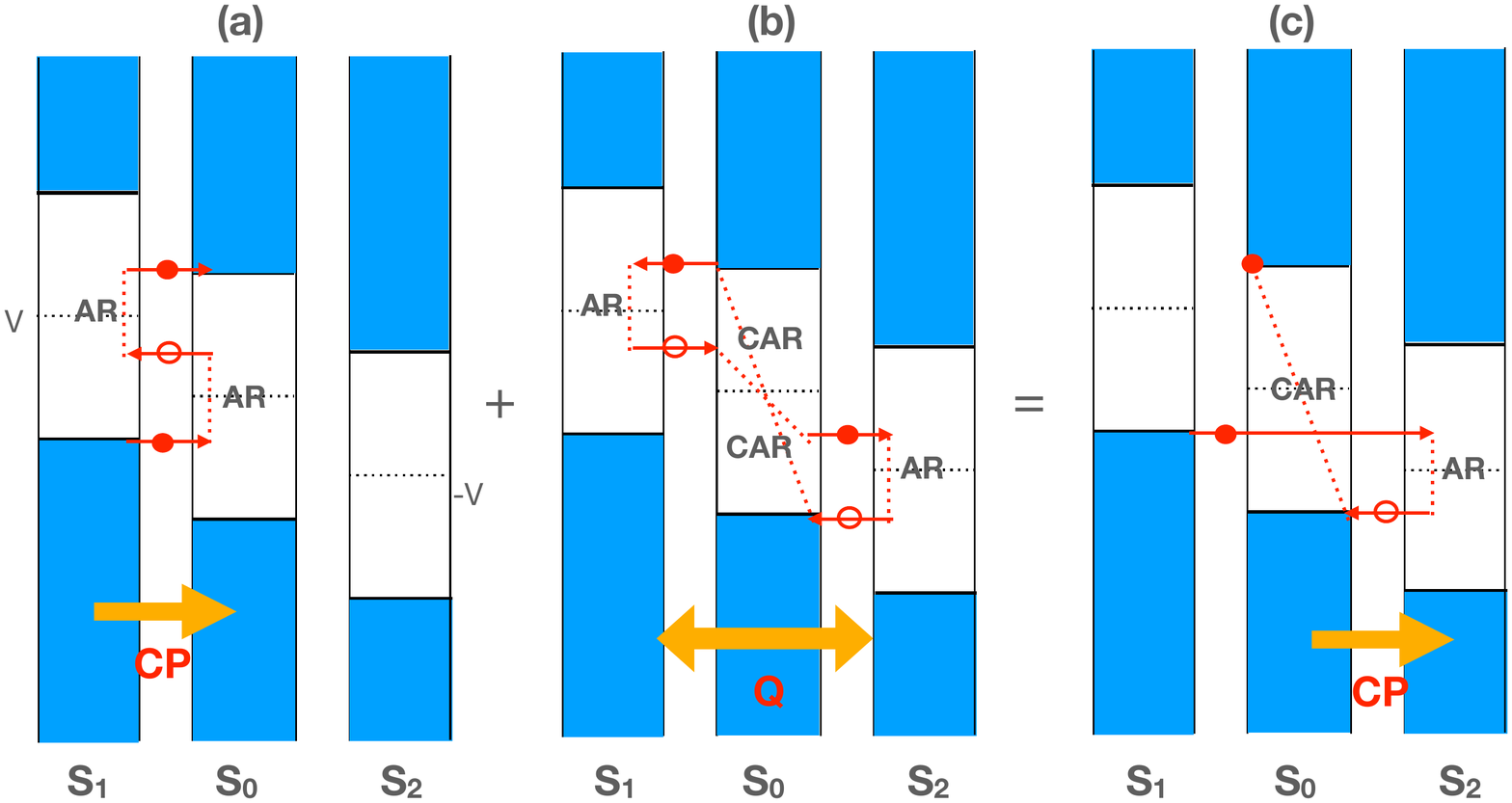}
  \end{minipage}\begin{minipage}{.39\textwidth}
    \caption{Diagram (a) pictures a quasiparticle promoted through the gap
      and a Cooper pair (CP) transferred from $S_1$ to $S_0$, via two
      Andreev reflections (AR). Diagram (b) pictures a quartet (Q) formed
      of two entangled Cooper pairs, transferred from $S_0$ to $S_1$ and
      $S_2$ simultaneously, with two AR and two crossed Andreev
      reflections (CAR). Diagram (c) pictures a quasiparticle promoted
      through the gap, transferred from $S_1$ to $S_0$, while a Cooper
      pair is instead transferred from $S_0$ to $S_2$. Diagram (c) can be
      formally obtained by superimposing the lines of diagrams (a) and
      (b), showing that phase-dependent MARs involve quartets.
      \label{fig:phaseMAR}}
  \end{minipage}
\end{figure*}

Besides this quartet supercurrent, another current component happens
to depend on the quartet phase. It originates from multiple Andreev
reflections (MAR), which promote quasiparticles across the
superconducting gap $2\Delta$ with the help of Cooper pair transfers,
each one gaining energy $2eV$ \cite{MAR}. New channels open in a
three-terminal Josephson junction (TTJ) \cite{Nowak}, where all pairs
of terminals are simultaneously involved. Among those, specific
processes involve emission of quartets at zero energy but with phase
$\varphi_Q$: the energy cost for promoting a quasiparticle between two
terminals, say $S_1$, $S_0$ (with $V_{1}-V_{0}=V$), instead of
transferring a pair between terminals $S_1$, $S_0$, can be provided by
transferring a pair between terminals $S_0$, $S_2$ (with
$V_{0}-V_{2}=V$) plus absorbing simultaneously a quartet from ($S_1$,
$S_2$) to $S_0$ (Figure \ref{fig:phaseMAR}). This quartet carries a
phase $\varphi_Q$ and these MAR processes become phase-dependent
subgap quasiparticle currents \cite{Jonckheere}. Detailed calculations
about the phase and voltage sensitivities of both quartet and
phase-MAR currents can be found in
Refs. \onlinecite{Sotto,Jonckheere2}.  While quartet supercurrents are
truly nondissipative, the phase-MAR currents are dissipative. Both of
them depend on the control variables $(\varphi_Q,V)$ but with
different symmetries \cite{Jonckheere,Sotto}. Owing to time inversion
symmetry, the quartet and phase-MAR currents have to be antisymmetric
with respect to inverting both variables $\varphi_Q$ and $V$. The
quartet current is antisymmetric in phase and symmetric in voltage,
but the phase-MAR current is instead symmetric in phase and
antisymmetric in voltage. This duality is reminiscent of the tunnel
junction treated by Josephson \cite{Josephson} in his seminal work,
concerning the DC current and the phase-sensitive quasiparticle
current. The latter is AC in a two-terminal junction, but can become
DC in a multiterminal one.

Regarding experiments, an important question is about the
interpretation of transport anomalies observed when a TTJ is biased at
the voltages $0,V,-V$
\cite{Lefloch,Heiblum,HGE,multiterminal-exp8,multiterminal-exp9,multiterminal-exp10}.
A conservative explanation involves the synchronization of AC
Josephson currents flowing across each of the junctions polarized at
$V$ and $-V$ respectively \cite{Jain}. This mechanism is
electromagnetic in nature and it involves the impedance (or the photon
modes) of the whole circuit including the junction.

Minimal models involve an adiabatic dependence of the currents with
time-dependent phases, in a way similar to the standard treatment of
Shapiro resonances \cite{Tinkham}. This can be done in the presence of
an external environment described by a circuit impedance, that
includes the resistive part of the junction itself, within RSJ-related
models \cite{Jain}. This qualitatively accounts for the DC-current
features observed in TTJs
\cite{multiterminal-exp3,multiterminal-exp2,multiterminal-exp4,multiterminal-exp8,multiterminal-exp5},
but is not a proof of the physical relevance of such a description. For
instance, the zero-frequency current-current cross-correlations
\cite{Heiblum} can hardly receive interpretation in terms of the RSJ
model, and quite specific frequency dependence of the device external
circuit impedance should be advocated to interpret a recent
four-terminal experiment as originating from a RSJ model
\cite{HGE}. Still, complementary experiments would be important to
ascertain the mesoscopic nature of multipair processes.

A first requisite is the control over the quartet phase, which can be
used to prove coherence of the multipair supercurrent. Such a phase
coherence might be present in an extrinsic synchronization scenario,
although hampered by decoherence mechanisms due to the environment
itself. On the contrary, phase coherence of the quartets is expected
to be much more robust.  To go further in the discrimination between
extrinsic and intrinsic mechanisms, one must take into account the
high transparency of the junctions, necessary to produce a mesoscopic
multipair transport. The consequence is the existence of MAR
processes, which in the standard two-terminal case have no explanation
but with the help of subgap Andreev reflections, and thus go well
beyond a phenomenological RSJ modeling. Specifically, in MTJs, the
observation of the phase-sensitive MARs can be taken as evidence
for truly mesoscopic processes involving quartets, thus disproving any
classical synchronization scenario.

In this work, we propose an interferometric scheme able to control the
quartet phase and, at the same time, reveal the phase-MAR component,
thus proving both the phase coherence of multipair processes and their
truly subgap mesoscopic nature.

Following Josephson's discovery that a current must flow in an
unbiased junction and depends on the phase difference between the
contacts \cite{Josephson}, SQUID setups were invented in order to
control and analyze this phase sensitivity \cite{Tinkham}. The flux
dependence exhibits period ${hc}/{2e}$ that directly proves
supercurrents carried by Cooper pairs with charge $2e$. Similarly, one
expects that interferometry also helps elucidating the mechanism of
quartets in TTJs, in particular proving that they carry a charge
4e. Yet, this simple expectation meets a difficulty : a TTJ involves
three terminals, two of them being biased. This prevents from building
a trivial generalization of the original two-terminal SQUID which is
fully equipotential. Such a device must necessarily be different from
those already proposed for multijunctions at equilibrium
\cite{multiterminal-exp1,biSQUID}.

In this work, we describe a four-terminal scheme building a true
quartet-SQUID. The clue is to connect two TTJs in parallel by their
unbiased as well as their biased terminals, in order to close them in
a double-TTJ loop. Cooper pairs injected in the quartet-SQUID at
voltage $V=0$ can cross either TTJ as quartets, picking up the quartet
phase of each TTJ, and recombine in the common outputs at voltages $V$
and $-V$. The design encloses two loops instead of one. Generalizing
the standard SQUID argument in the presence of magnetic flux shows
that this imposes a difference between the quartet phases of the two
TTJs, thus achieving a perfect parallel with an ordinary SQUID.

This scheme allows analyzing the sensitivity of the quartet mode on
voltage, as a new control parameter for a DC supercurrent. Microscopic
models show that it is not monotonous, owing to nonadiabatic
transitions between Andreev levels. Moreover it can switch from a
generic $\pi$-junction behavior (perturbative and low voltage case) to
a $0$-junction one. Such an evidence goes beyond classical
synchronization scenarii unless assuming ad hoc an unlikely voltage
(i.e. AC Josephson frequency) dependence of the circuit impedance.

The proposed quartet-SQUID also allows exploiting the interplay
between quartets and phase-sensitive MARs.  Separation between those
two distinct processes could in principle be achieved in ideally
symmetrical TJJs. More generally, the different phase symmetry of
quartet and phase-sensitive MAR currents results in a phase lapse in
the periodic flux response of the quartet-SQUID. Measuring this phase
lapse quantifies the presence of phase-MARs in transparent enough
junctions. Phase-MARs are mesoscopic and they involve quartet
excitation amplitudes, therefore they bring the necessary proof of a
truly new physics being involved in TTJs.

\begin{figure*}[t]
  \begin{minipage}{.6\textwidth}
    \includegraphics[width=1\textwidth]{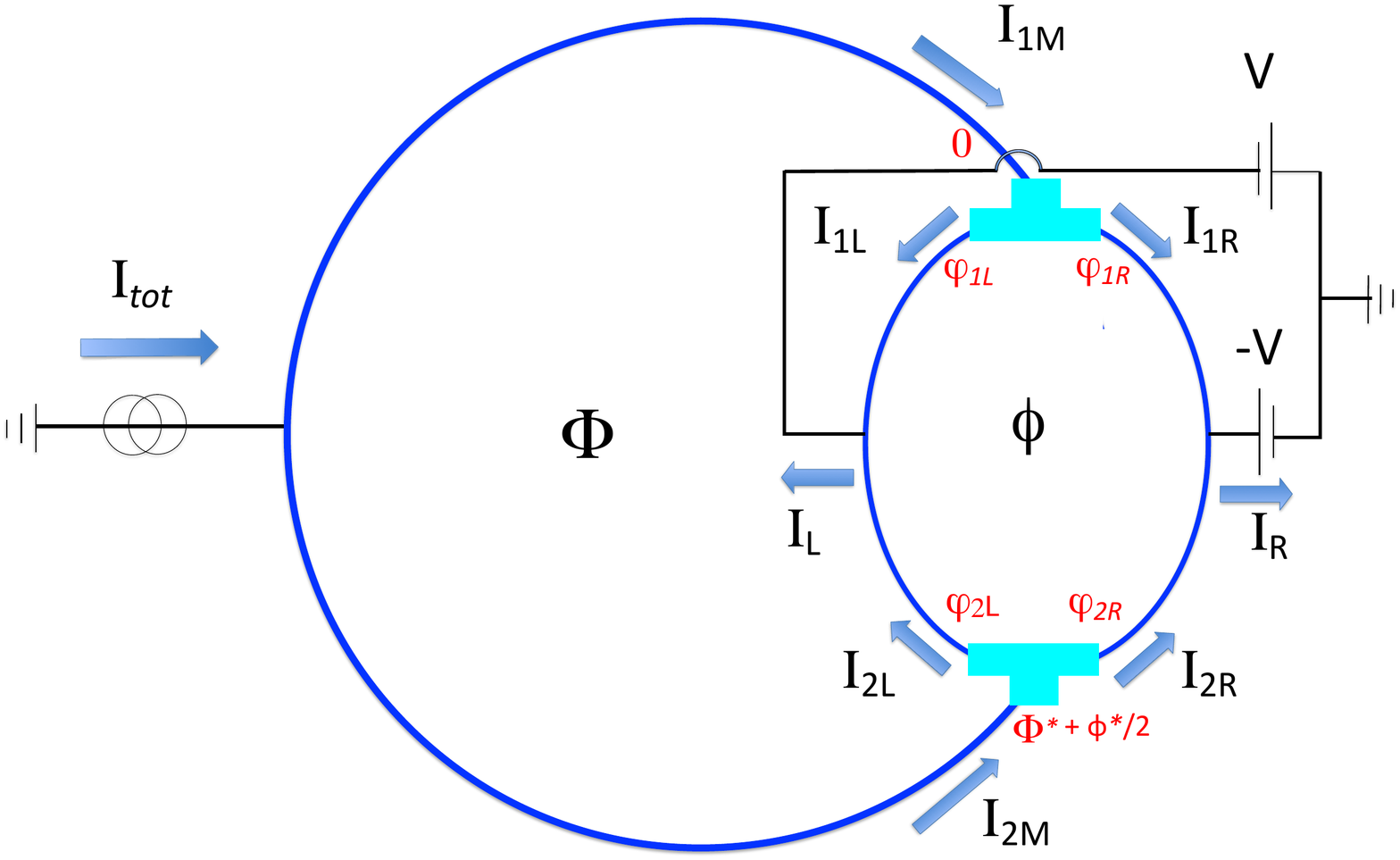}
  \end{minipage}\begin{minipage}{.39\textwidth}
    \caption{Scheme of a quartet-SQUID based on two TTJs, with a main loop
      threaded by a flux $\Phi$ and a secondary loop threaded by a flux
      $\phi$. The current is injected into the large loop such as to make
      the two TTJs interfere with each other. The current exits through
      the biased leads at voltages $V, -V$ of the branches $L,R$ of the
      secondary loop. The phases are mentioned in red within a simple
      gauge convention, see text. Here $\varphi_{1M}=0$ and
      $\varphi_{2M}=\Phi^*+{\phi^*}/{2}$.
      \label{QuartetSQUID}
    }
  \end{minipage}
\end{figure*}

{\it Three-terminal junction quartet-SQUID:} The principle of the
quartet-SQUID is to make two TTJs interfere with each other by joining
their biased arms in a secondary circuit, as pictured in Figure
\ref{QuartetSQUID}. The two TTJs thus enclose a secondary loop with
two branches respectively at the voltages $V$ (hereafter denoted as
``$L$-branch''), $-V$ (hereafter denoted as ``$R$-branch''), threaded
by a flux $\phi$. The main loop is threaded by a flux $\Phi$. Both
loops are separated by the $L$ branch (see
Figure~\ref{QuartetSQUID}). The total current is injected as
$I_{tot}=I_{1M}+I_{2M}$ where $I_{1M}$ and $I_{2M}$ are the currents
entering each of the TTJs from the unbiased branch, and eventually
exiting in the biased branches (second circuit) as $I_L$ and
$I_R$. Current conservation reads:
\begin{equation}
  \label{totalcurrent}
  I_{tot}=I_{1M}+I_{2M}=I_L+I_R
\end{equation}

Let us define the phases at the unbiased branch of TTJ$1$ and TTJ$2$
as $\varphi_{1M},\varphi_{2M}$ respectively, and the phases at the
biased branches of the TTJs as $(\varphi_{1L},\varphi_{1R})$ and
$(\varphi_{2L},\varphi_{2R})$ respectively. From previous works \cite{Freyn} one
knows that the stationary quartet phase components are
\begin{equation}
  \label{quartetphase}
  \varphi_{Qi}=\varphi_{iL}+\varphi_{iR}-2\varphi_{iM},
\end{equation} 
while the oscillating phase components (at frequency ${4eV}/{\hbar}$)
are $\varphi_{iL}-\varphi_{iR}$ ($i=1,2$).  Let us define the
normalized fluxes between $0$ and $2\pi$ as
$\Phi^*=({2\pi}/{\phi_0})\Phi$ and $\phi^*=({2\pi}/{\phi_0})\phi$,
with $\phi_0={hc}/{2e}$. The fluxoid argument is applied to the main
loop containing the $L$ branch, then to the main plus secondary loop
containing the $R$ branch. This is perfectly allowed, in spite of the
main loop and the biased branches not being at the same potential. In
fact, the fluxoid argument takes care of the phase variation inside
each superconductor, whatever its potential. The supercurrent
circulation in the bulk of each superconductor is assumed to be zero
as for a thick superconductor, see Ref.~\onlinecite{Tinkham}. The
presence of voltage biases between the different superconductors only
enters in the phase difference at the junctions, that can depend on
time in the present scheme, with frequency ${2eV}/{\hbar}$. The
fluxoid argument \cite{Tinkham} amounts to equating on both paths the
sum of the phase differences at the junctions to the normalized flux
in the loop (modulo $2\pi$), which yields:
\begin{eqnarray}
  \label{fluxoid-1}
  \Phi^*&=&\varphi_{1L}-\varphi_{1M}+\varphi_{2M}-\varphi_{2L}\\
  \Phi^*+\phi^*&=&\varphi_{1R}-\varphi_{1M}+\varphi_{2M}-\varphi_{2R}
  \label{fluxoid-2}
  .
\end{eqnarray}
Taking the difference between these two equations, one obtains a relation
between the oscillating phases components at the two TTJs:
\begin{equation}
  (\varphi_{1R}-\varphi_{1L})-(\varphi_{2R}-\varphi_{2L})=\phi^*
  ,
\end{equation}
expressing that these time-dependent components are perfectly
synchronized. On the other hand, taking the sum of
Eqs.~(\ref{fluxoid-1})-(\ref{fluxoid-2}) yields a relation between the
quartet phases of the two TTJs [see Eq.~(\ref{quartetphase})]:
\begin{equation}
  \label{ddphasequartet}
  \varphi_{1Q}-\varphi_{2Q}=2(\Phi^*+{\phi^*}/{2})
\end{equation}

This central result shows that, like an ordinary SQUID, the
interferometer imposes a phase difference between the stationary
quartet phases at the two TTJs. Because of the ($L$, $R$) symmetry of
the quartet current, the corresponding flux is the arithmetic mean of
the fluxes delimited by the $L$ (i.e. $\Phi^*$) and the $R$
(i.e. $\Phi^*+\phi^*$) branches.

Interestingly, if the TTJs are symmetric by exchanging their contacts
to branches $L,R$, the currents $I_{1,2M}$ entering the TTJs are pure
quartet currents
i.e. $I_{1M}=I_{1Q}(\varphi_{1Q}),I_{2M}=I_{2Q}(\varphi_{2Q})$. In
turn, a pure MAR current $I_L-I_R$ flows between branches $L$ and $R$
thus in the secondary circuit, and one can write:
\begin{eqnarray}
  I_L&=&\frac{1}{2}\big[I_{1Q}(\varphi_{1Q})+I_{2Q}(\varphi_{2Q})\big]\\
  \nonumber&+&I_{1MAR}(\varphi_{1Q})+I_{2MAR}(\varphi_{2Q})\\
  I_R&=&\frac{1}{2}\big[I_{1Q}(\varphi_{1Q})+I_{2Q}(\varphi_{2Q})\big]\\
  \nonumber&-&I_{1MAR}(\varphi_{1Q})-I_{2MAR}(\varphi_{2Q})
  .
\end{eqnarray}
In this case, a Cooper pair current circulates in the main loop, while
the secondary one contains a superposition of a quartet current --
flowing as parallel Cooper pair currents in the $L$ and $R$ branches,
thus insensitive to the flux $\phi^*$ -- and a circulating MAR current
-- sensitive to $\phi^*$ via its phase-MAR component.

In the general case of asymmetric TTJs, all currents
$I_{1,2M},I_{L,R}$ contain components of both quartet and MAR
origins. One can carry the analysis further in the simplifying case of
weak transparencies. Noting that the quartet and MAR components are
respectively odd and even in the quartet phases, and one can write:
\begin{eqnarray}
  \label{tunnelcurrents-1}
  I_{1M}&=&I_{Qc1}(V)\sin\varphi_{1Q}+\bar{I}_{MAR1}(V)+I_{MARc1}(V)\cos\varphi_{1Q}\\
  \label{tunnelcurrents-2}
  I_{2M}&=&I_{Qc2}(V)\sin\varphi_{2Q}+\bar{I}_{MAR2}(V)+I_{MARc2}(V)\cos\varphi_{2Q}
  ,
\end{eqnarray}
where the first terms in
Eqs.~(\ref{tunnelcurrents-1})-(\ref{tunnelcurrents-2}) are the quartet
currents, with ``critical currents'' $I_{Qci}$. The second terms are
the phase-averaged MAR components, including the phase-independent
two-terminal MAR processes, and the last terms contain the
phase-sensitive MAR components, with ``critical currents''
$I_{MARci}$. The critical current values defining the amplitude of the
phase oscillations of the quartet and MAR currents are
voltage-sensitive, and have in general nonmonotonous variations with
$V$ \cite{Sotto,paperI,paperII,PRB2022,Jonckheere2}. The sine and cosine
dependences of the respective quartet and MAR currents stem from their
symmetry in phase. Such expressions can be checked by microscopic
calculations in the low transparency case \cite{Sotto}.

From Eqs. (\ref{totalcurrent}), (\ref{ddphasequartet}),
(\ref{tunnelcurrents-1}) and~(\ref{tunnelcurrents-2}), the total
current injected in this quartet-SQUID can be written as (omitting the
voltage sensitivities):
\begin{eqnarray}
  I_{tot}&=&\bar{I}_{MAR1}+\bar{I}_{MAR2}+I_{c1}\sin(\varphi_{1Q}+\alpha_1)\\
  &+&I_{c2}\sin\big(\varphi_{1Q}-2(\Phi^*+{\phi^*}/{2})+\alpha_2\big)
  ,
\end{eqnarray}
with ($i=1,2$):
\begin{eqnarray}
  \nonumber
  I_{ci}&=&\sqrt{I_{Qci}^2+I_{MARci}^2}\\
  \tan(\alpha_i)&=&{I_{MARci}}/{I_{Qci}}
  .
\end{eqnarray}

The total current appears as the sum of (i) a phase-independent MAR
current and (ii) a typical SQUID current, which depends on the quartet
phase $\varphi_{1Q}$, and on the effective flux $\Phi^*+{\phi^*}/{2}$,
with phase lapses $\alpha_{1,2}$ that measure the ratio of
phase-sensitive MAR currents to quartet currents. As in a usual SQUID,
maximizing the total current with respect to the (quartet) phase
yields the following expression for the critical current:
\begin{eqnarray}
  \label{Itotal}
  I_{tot}&=&\bar{I}_{MAR1}+\bar{I}_{MAR2}\\
  \nonumber
  &+&\Big[I_{c1}^2+I_{c2}^2+2I_{c1}I_{c2}\cos\big(2(\Phi^*+{\phi^*}/{2})+\alpha_1-\alpha_2\big)\big]^{1/2}
  .
\end{eqnarray}

This relation achieves the goal of building a quartet-SQUID. As a
first result, the factor 2 in the flux sensitivity, that results in a
${hc}/{4e}$ periodicity, manifests the fact that quartets are made of
two entangled Cooper pairs and carry charge $4e$. Second, the phase
lapses $\alpha_{1,2}$ directly contain the information about the
presence or not of phase-MARs. These phase lapses disappear in the
case of TTJs with symmetric branches $V,-V$ ($\alpha_{1,2}=0$) or in
the unlikely case of identical TTJs ($\alpha_1=\alpha_2$).

In experiments performed at low voltage and in incoherent diffusive
regimes, the MAR currents are negligible, and the quartet-SQUID gives
direct access to the pure quartet currents.

The above discussion is not restricted to harmonic dependences of the
quartet and MAR current with phase. In resonant dot models,
nonharmonic behavior is easily obtained and the quartet current can be
quite large, actually comparable to the ordinary Josephson current of
a two-terminal junction in the same conditions
\cite{Jonckheere,Sotto}.

{\it Exploring the voltage dependence: from ``$0-$'' to ``$\pi-$''
  junction:} Having a quartet-SQUID in hands allows a thorough study
of the dependence of the quartet (and phase-MAR) currents with
voltage, as a new control parameter for DC Josephson currents.
Focusing on the quartet current, different models, suited to different
types of junctions (single or many level quantum dot, or diffusive
metallic) lead to the same conclusions: the quartet current-phase
characteristics changes sign several times with voltage, owing to
nonadiabatic transitions between Andreev levels, triggered by the
voltage via the running phase $(\varphi_L-\varphi_R)(t)$ at frequency
${4eV}/{\hbar}$ \cite{Sotto,paperI,paperII,PRB2022,Jonckheere2}.  This means
that, in terms of the quartet current component, a TTJ can be either a
``$0-$'' or a ``$\pi-$'' junction, with respect to the quartet phase
$\varphi_Q$. The same occurs with the phase-MAR current component that
also changes sign but at different voltages. Superposition of quartet
and phase-MAR components actually makes a generic TTJ a
``$\theta$-junction''.

More generally, the characteristics of a TTJ (transparency,
asymmetries between the three contacts, degree of decoherence) all
conspire to shift or even suppress the sign changes.  For instance, if
the couplings to the biased terminals are much smaller than the one to
the unbiased terminal, and the quartet TTJ current keeps a ``$\pi-$''
junction character at low and intermediate voltages
\cite{Jonckheere2}. Focusing on quartets only, in the case where
backgates allow to separately control the transparency of the
different contacts, one can reach a situation where, for a given
voltage, the pair of TTJs of the SQUID can be both ``$0-$'' (or both
``$\pi-$'') junctions, or one being a ``$0-$'' and the other a
``$\pi-$'' junction. This strongly reminds the experiments performed
with carbon nanotubes \cite{Wernsdorfer} (nanoSQUID) where the
mechanism for ``$0-$'' to ``$\pi-$'' transition is instead the Coulomb
interaction and the gate control of the nanotube junctions. In
addition, ``$0-$'' to ``$\pi-$'' transitions have also been observed
in superconductor-ferromagnet-superconductor Josephson junctions
\cite{SFS}.

To illustrate the possibilities of such a quartet-SQUID, let us assume
that TTJ1 is fully symmetric and resonant, with high quartet critical
currents and several sign changes as $V$ is increased from $0$ to
$2\Delta$. On the contrary, TTJ2 couples weakly but equally to the
$L$, $R$ terminals. This suppresses the MAR component in the SQUID
current, and leaves us with a very asymmetric quartet-SQUID, with
(neglecting the anharmonicity in this example):

\begin{equation}
  I_{tot}(V)=I_{c1}(V)\sin(\varphi_{1Q})
  +I_{c2}(V)\sin\big(\varphi_{1Q}-2(\Phi^*+{\phi^*}/{2})\big)
  \label{asymSQUID}
\end{equation}
and $|I_{c1}(V)|>>|I_{c2}(V)|$. As said above, TTJ2 remains a ``$\pi-$''
junction so that $I_{c2}<0$, while the sign of $I_{c1}$ depends on
$V$. Following the classical argument of an asymmetric SQUID, one first
maximizes $I_{tot}\sim I_{c1}\sin(\varphi_{1Q})$ with respect to
$\varphi_{1Q}$, which yields $\varphi_{1Q}\sim {\pi}/{2}$ if
$I_{c1}>0$ and $\varphi_{1Q}\sim3{\pi}/{2}$ if $I_{c1}<0$. Inserting
this value into the -- small -- second term of Eq.~(\ref{asymSQUID})
yields

\begin{equation}
  I_{tot}\sim|I_{c1}|
  \pm I_{c2}\cos2(\Phi^*+{\phi^*}/{2})
  ,
\end{equation}
with $\pm$ sign depending on the ``$0-$'' or ``$\pi-$'' character of
TTJ1. First, this reconstructs the current-phase relation of
TTJ2, including the sign of $I_{c2}$. Second, as $V$ is swept upwards
from $0$, the sign changes of TTJ1 reflect themselves in $\pi-$ shifts
in the flux dependence of $I_{tot}$.

As another example, Eq.~(\ref{Itotal}) shows that phase-MARs can be
investigated in TTJ1 only, provided TTJ2 is symmetric in ($L$, $R$)
thus $\alpha_2=0$. The relative amplitude of phase-MARs and quartets
in TTJ1 reflects directly in the shift $\alpha_1$ of the total current
vs flux dependence. In the generic case of non-symmetric TTJs, one
instead measures the difference $\alpha_1-\alpha_2$ of the two TTJ
phase lapses.

An additional piece of information can also be gained by measuring the
currents in the biased loop, i. e. the combination $I_L-I_R$ which,
contrarily to $I_{tot}$, eliminates the quartet components at low
transparency  and is thus sensitive to MAR currents
only.

{\it Conclusion:} We have proposed a quartet-SQUID
generalizing the standard SQUID geometry to make quartet and
phase-sensitive MAR current interfere under control of a magnetic
flux. The periodicity in the flux dependence of the total critical
current through the SQUID reflects the quartet charge $4e$. In
addition, the distinguishing phase symmetries of both current
components imply a phase lapse in the flux sensitivity of the critical
current of the interferometer, which allows to quantify the phase-MARs
with respect to the quartet current. Finally, phase-MARs are a
consequence of both quartet emission and coherent subgap
transport. Thus, they provide evidence against scenarii based on
extrinsic synchronization via the outer circuit of junctions described
by an adiabatic current-phase relation. A full quantitative analysis and comparison with future experiments requires to inject into the present description the expressions for quartet and MAR currents of each TTJ, obtained from microscopic theories.The principle of the present
quartet-SQUID can obviously be generalized to higher order multipair
transport in MTJs with four or more terminals. 

R.M. acknowledges support from the French National Research Agency
(ANR) in the framework of the Graphmon project (ANR-19-CE47-0007).


\begin{thebibliography}{99}

\bibitem{Omelyanchouk1} R. de Bruyn Ouboter and A. Omelyanchouk,
  {Multi-terminal squid controlled by the transport current},
  Physica B: Condensed Matter {\bf 205}, 153 (1995).

\bibitem{Omelyanchouk2} M. Amin, A. Omelyanchouk, and A. Zagoskin,
  {Mesoscopic multiterminal Josephson structures. i. effects of
    nonlocal weak coupling}, Low Temperature Physics {\bf 27}, 616 (2001).

\bibitem{Omelyanchouk3} M. Amin, A. Omelyanchouk, and A. Zagoskin, {Dc
  squid based on the mesoscopic multiterminal Josephson junction},
  Physica C: Superconductivity {\bf 372}, 178 (2002).

\bibitem{Omelyanchouk4} M. Amin, A. Omelyanchouk, A. Blais, A. M. van
  den Brink, G. Rose, T. Duty, and A. Zagoskin, {Multi-terminal
    superconducting phase qubit}, Physica C: Superconductivity {\bf
    368}, 310 (2002).

\bibitem{vanHeck} B. van Heck, S. Mi, and A.R. Akhmerov, {Single
  fermion manipulation via superconducting phase differences in
  multiterminal Josephson junctions}, Phys.  Rev. B \textbf{90},
  155450 (2014).

\bibitem{Padurariu} C. Padurariu, T. Jonckheere, J. Rech,
  R. M\'{e}lin, D. Feinberg, T. Martin, and Yu.V. Nazarov,
  {Closing the proximity gap in a metallic Josephson junction
    between three superconductors}, Phys. Rev. B \textbf{92}, 205409
  (2015).

\bibitem{Nazarov1} R.-P. Riwar, M. Houzet, J.S. Meyer, and
  Y.V. Nazarov, {Multi-terminal Josephson junctions as topological
    materials}, Nat. Commun. {\bf 7}, 11167 (2016).
  
\bibitem{Nazarov2} E. Eriksson, R.-P. Riwar, M. Houzet, J. S. Meyer,
  and Y. V. Nazarov, {Topological transconductance quantization in
    a four-terminal Josephson junction}, Phys. Rev. B {\bf 95}, 075417
  (2017).

\bibitem{Levchenko1} H.-Y. Xie, M.G. Vavilov and A. Levchenko, {
  Topological Andreev bands in three-terminal Josephson junctions},
  Phys. Rev. B {\bf 96}, 161406 (2017).

\bibitem{Levchenko2} H.-Y. Xie, M.G. Vavilov and A. Levchenko, {
  Weyl nodes in Andreev spectra of multiterminal Josephson junctions:
  Chern numbers, conductances and supercurrents}, Phys. Rev. B {\bf
  97}, 035443 (2018).
  
\bibitem{topo0} O. Deb, K. Sengupta and D. Sen, {Josephson
  junctions of multiple superconducting wires}, Phys. Rev. B 97,
  174518 (2018).

\bibitem{Feinberg1} B. Venitucci, D. Feinberg, R. M\'elin,
  B. Dou\c{c}ot, {Nonadiabatic Josephson current pumping by
    microwave irradiation}, Phys. Rev. B {\bf 97}, 195423 (2018).

\bibitem{Feinberg2} L.P. Gavensky, G. Usaj, D. Feinberg and
  C.A. Balseiro, {Berry curvature tomography and realization of
    topological Haldane model in driven three-terminal Josephson
    junctions}, Phys. Rev. B {\bf 97}, 220505 (2018).

\bibitem{topo1-plus-Floquet} R. L. Klees, G. Rastelli, J. C. Cuevas,
  and W. Belzig, {Microwave Spectroscopy Reveals the Quantum Geometric
    Tensor of Topological Josephson Matter}, Phys. Rev. Lett. {\bf
    124}, 197002 (2020).
  
\bibitem{Berry} B. Dou\c{c}ot, R. Danneau, K. Yang, J.-G. Caputo and
  R. M\'elin, {Berry phase in superconducting multiterminal
    quantum dots}, Phys. Rev. B {\bf 101}, 035411 (2020).
  
\bibitem{topo2} V. Fatemi, A.R. Akhmerov and L. Bretheau,
  {Weyl Josepshon circuits}, Phys. Rev. Research 3, 013288 (2021).

\bibitem{topo3} L. Peyruchat, J. Griesmar, J.-D. Pillet and \c{C}.\"O
  Girit, {Transconductance quantization in a topological Josephson
    tunnel junction circuit}, Phys. Rev. Research 3, 013289 (2021).
  
\bibitem{topo1} H. Weisbrich, R.L. Klees, G. Rastelli and W. Belzig,
  {Second Chern Number and Non-Abelian Berry Phase in Topological
    Superconducting Systems}, PRX Quantum {\bf 2}, 010310 (2021).

\bibitem{topo4} Y. Chen and Y.V. Nazarov, {Weyl point immersed in
  a continuous spectrum: an example from superconducting
  nanostructures}, Phys. Rev. B {\bf 104}, 104506 (2021).

\bibitem{topo4-bis} Y. Chen and Y.V. Nazarov, {Spin-Weyl quantum
  unit: theoretical proposal}, Phys. Rev. B {\bf 103}, 045410 (2021).
  
\bibitem{topo5} E.V. Repin and Y.V. Nazarov, {Weyl points in the
  multi-terminal Hybrid Superconductor-Semiconductor Nanowire
  devices}, Phys. Rev. B {\bf 105}, L041405 (2022)

\bibitem{Akh2} A. Melo, V. Fatemi and A.R. Akhmerov, {Multiplet
  supercurrent in Josephson tunneling circuits}, SciPost Phys. {\bf
  12}, 017 (2022)

\bibitem{Gavensky} L. Peralta Gavensky, G. Usaj and C.A. Balseiro,
  {Multiterminal Josephson junctions: a road to topological flux
    networks}, arXiv:2211.12524 (2022).
  
\bibitem{Cuevas-Pothier} J. C. Cuevas and H. Pothier, {
  Voltage-induced Shapiro steps in a superconducting multiterminal
  structure}, Phys. Rev. B {\bf 75}, 174513 (2007).

\bibitem{Houzet-Samuelsson} M. Houzet and P. Samuelsson, Multiple
  Andreev reflections in hybrid multiterminal junctions, Phys. Rev. B
  {\bf 82}, 060517(R) (2010).
  
\bibitem{Freyn} A. Freyn, B. Dou\c{c}ot, D. Feinberg, and R. M\'elin,
  {Production of non-local quartets and phase-sensitive
    entanglement in a superconducting beam splitter},
  Phys. Rev. Lett. {\bf 106}, 257005 (2011).

\bibitem{Jonckheere}  T. Jonckheere, J. Rech, T. Martin,
  B. Dou\c{c}ot, D. Feinberg, and R. M\'{e}lin, {Multipair DC
    Josephson resonances in a biased allsuperconducting bijunction},
  Phys. Rev. B \textbf{87}, 214501 (2013).
  
\bibitem{biSQUID} J. Rech, T. Jonckheere, T. Martin, B. Dou\c{c}ot,
  D. Feinberg and R. M\'elin, Phys. Rev. B {\bf 90}, 075419 (2014).

\bibitem{EPJB} D. Feinberg, T. Jonckheere, J. Rech, T. Martin,
  B. Dou\c{c}ot and R. M\'elin, Quartets and the current-phase
  structure of a double quantum dot superconducting bijunction at
  equilibrium, Eur. Phys. J. B {\bf 88}, 99 (2015).
  
\bibitem{Melin1} R. M\'elin, D. Feinberg, and B. Dou\c{c}ot, {
  Partially resummed perturbation theory for multiple Andreev
  reflections in a short three-terminal Josephson junction},
  Eur. Phys. J. B {\bf 89}, 67 (2016). 

\bibitem{Sotto} R. M\'elin, M. Sotto, D. Feinberg,
  J.-G. Caputo and B. Dou\c{c}ot, {Gate-tunable zero-frequency
    current cross-correlations of the quartet mode in a voltage-biased
    three-terminal Josephson junction}, Phys. Rev. B {\bf 93}, 115436
  (2016).

\bibitem{FWS} R. M\'elin, J.-G. Caputo, K. Yang and B. Dou\c{c}ot,
  {Simple Floquet-Wannier-Stark-Andreev viewpoint and emergence of
    low-energy scales in a voltage-biased three-terminal Josephson
    junction}, Phys. Rev. B {\bf 95}, 085415 (2017).

\bibitem{engineering} R. M\'elin, R. Danneau,
  K. Yang, J.-G. Caputo, and B. Dou\c{c}ot, {Engineering the
    Floquet spectrum of superconducting multiterminal quantum dots},
  Phys. Rev. B {\bf 100}, 035450 (2019).

\bibitem{Nowak} Nowak, M. P., Wimmer, M., Akhmerov, A., {Supercurrent
  carried by nonequilibrium quasiparticles in a multiterminal
  Josephson junction}, Phys. Rev. B {\bf 99}, 075516 (2019).

\bibitem{Pillet} J.D. Pillet, V. Benzoni, J. Griesmar, J.-L. Smirr,
  and \c{C}.\"{O}. Girit, {Nonlocal Josephson Effect in Andreev
    Molecules} Nano Lett. {\bf 19}, 7138 (2019).

\bibitem{Nazarov-PRR} V. Kornich, H.S. Barakov, and Yu.V. Nazarov,
  {Fine energy splitting of overlapping Andreev bound states in
    multiterminal superconducting nanostructures}, Phys. Rev. Research
  {\bf 1}, 033004 (2019).
  
\bibitem{paperI} R. M\'elin, {Inversion in a four terminal
  superconducting device on the quartet line. I. Two-dimensional metal
  and the quartet beam splitter}, Phys. Rev. B {\bf 102}, 245435
  (2020).

\bibitem{paperII} R. M\'elin and B. Dou\c{c}ot, {Inversion in a
  four terminal superconducting device on the quartet
  line. II. Quantum dot and Floquet theory}, Phys. Rev. B {\bf 102},
  245436 (2020).

\bibitem{Pillet2} J.-D. Pillet, V. Benzoni, J. Griesmar, J.-L. Smirr,
  and \c{C} \"O Girit, {Scattering description of Andreev
    molecules}, SciPost Phys. Core {\bf 2}, 009 (2020).

\bibitem{Nazarov-PRB-AM} V. Kornich, H. S. Barakov and Yu. V. Nazarov,
  {Overlapping Andreev states in semiconducting nanowires:
    competition of 1D and 3D propagation}, Phys. Rev. B {\bf 101},
  195430 (2020).
  
\bibitem{long-distance} R. M\'elin, {Ultralong-distance quantum
  correlations in three-terminal Josephson junctions}, Phys. Rev. B
  {\bf 104}, 075402 (2021).

\bibitem{PRB2022} R. M\'elin, Multiterminal ballistic Josephson
  junctions coupled to normal leads, Phys. Rev. B {\bf 105}, 155418
  (2022).
  
\bibitem{Lefloch} A.H. Pfeffer, J.E. Duvauchelle, H. Courtois,
  R. M\'elin, D. Feinberg, and F. Lefloch, {Subgap structure in the
    conductance of a three-terminal Josephson junction}, Phys. Rev. B
  {\bf 90}, 075401 (2014).
  
\bibitem{multiterminal-exp1}  E. Strambini, S. D'Ambrosio, F. Vischi,
  F.S. Bergeret, Yu.V. Nazarov, and F. Giazotto, {The
    $\omega$-SQUIPT as a tool to phase-engineer Josephson topological
    materials}, Nat. Nanotechnol. \textbf{11}, 1055 (2016).
  
\bibitem{Heiblum} Y. Cohen, Y. Ronen, J.H. Kang, M. Heiblum,
  D. Feinberg, R.  M\'elin, and H. Strikman, {Non-local
    supercurrent of quartets in a three-terminal Josephson junction},
  Proc. Natl. Acad. Sci. U. S. A. \textbf{115}, 6991 (2018).
  
\bibitem{multiterminal-exp2} A.W. Draelos, M.-T. Wei, A. Seredinski, H. Li,
  Y. Mehta, K. Watanabe, T. Taniguchi, I.V. Borzenets, F. Amet, and
  G. Finkelstein, {Supercurrent flow in multiterminal graphene
    Josephson junctions}, Nano Lett. \textbf{19}, 1039 (2019).  
  
\bibitem{HGE} K.F. Huang, Y. Ronen, R. M\'elin, D. Feinberg,
  K. Watanabe, T. Taniguchi, and P. Kim, {Evidence for 4e charge of
    Cooper quartets in a biased multi-terminal graphene-based
    Josephson junction}, Nat. Comm. {\bf 13}, 3032 (2022).
  
\bibitem{multiterminal-exp3} N.  Pankratova, H.  Lee, R. Kuzmin, K.
  Wickramasinghe, W. Mayer,J. Yuan,M. Vavilov,J. Shabani and
  V. Manucharyan, {The multi-terminal Josephson effect},
  Phys. Rev. X {\bf 10}, 031051 (2020).

\bibitem{multiterminal-exp4} G.V. Graziano, J.S. Lee, M. Pendharkar,
  C. Palmstrom and V.S. Pribiag, {Transport Studies in a
    Gate-Tunable Three-Terminal Josephson Junction}, Phys. Rev. B {\bf
    101}, 054510 (2020).
  
\bibitem{multiterminal-exp5} E.G. Arnault, T. Larson, A. Seredinski,
  L. Zhao, H. Li, K. Watanabe, T. Taniguchi, I. Borzenets, F. Amet
  and G. Finkelstein, {The multiterminal inverse AC Josephson
    effect}, Nano Lett. {\bf 21}, 9668 (2021).

\bibitem{multiterminal-exp6} S.A. Khan, L. Stampfer, T. Mutas,
  J.-H. Kang, P. Krogstrup and T.S. Jespersen, {Multiterminal
    Quantized Conductance in InSb Nanocrosses}, Advanced Materials
  {\bf 33}, 2100078 (2021).

\bibitem{multiterminal-exp7} O. K\"urt\"ossy, Z. Scher\"ubl,
  G. F\"ul\"op, I. E. Luk\'acs, T. Kanne, J. Nygard, P. Makk and
  S. Csonka, {Andreev molecule in parallel InAs nanowires}, Nano
  Lett. {\bf 21}, 7929 (2021).

\bibitem{multiterminal-exp8} G. V. Graziano, M. Gupta, M. Pendharkar, J. T. Dong,
  C. P. Dempsey, C. Palmstr{\o}m and V. S. Pribiag, Selective control
  of conductance modes in multi-terminal Josephson junctions, Nat.
  Comm. {\bf 13}, 5933 (2022).

\bibitem{multiterminal-exp9} E.G. Arnault, S. Idris, A. McConnell,
  L. Zhao, T.F.Q. Larson, K. Watanabe, T. Taniguchi, G. Finkelstein,
  F. Amet, {Dynamical stabilization of multiplet supercurrents in
    multi-terminal Josephson junctions}, Nano Lett. {\bf 22}, 7073
  (2022).

\bibitem{multiterminal-exp10} F. Zhang, A.S. Rashid, M.T. Ahari,
  W. Zhang, K.M. Ananthanarayanan, R. Xiao, G.J. de Coster,
  M.J. Gilbert, N. Samarth and M. Kayyalha, {Andreev processes in
    mesoscopic multi-terminal graphene Josephson junctions}, arXiv:
  2210.04408v2 (2022).
  
\bibitem{MAR} T. M. Klapwjik, G. E. Blonder, and M. Tinkham,
  Explanation of subharmonic energy gap structure in superconducting
  contacts, Physica B+C 109-110, 1657 (1982); M. Octavio, M. Tinkham,
  G. E. Blonder, and T. M. Klapwijk, Subharmonic energy-gap structure
  in superconducting constrictions, Phys. Rev. B 27, 6739 (1983).

\bibitem{Jonckheere2} T. Jonckheere, J. Rech, C. Padurariu,
  L. Raymond, T. Martin, D. Feinberg, in preparation.

\bibitem{Josephson} B.D. Josephson, Possible new effects in
  superconductive tunnelling, Physics Letters 1, 251 (1962).

\bibitem{Jain} Jain, A. K., Likharev, K. K., Lukens J. E., and
  Sauvageau, J. E., Mutual Phase-locking in Josephson junction arrays.
  Phys. Rep. 109, 309-426 (1984).

\bibitem{Tinkham} M. Tinkham, Introduction to Superconductivity, 2nd
  ed.  (McGraw-Hill, New York, 1996).
  
\bibitem{Wernsdorfer} J.-P. Cleuziou, W. Wernsdorfer, V. Bouchiat, T.
  Ondarc\c{c}uhu, and M. Monthioux, {Carbon Nanotube Superconducting
    Quantum Interference Device}, Nature Nanotech. {\bf 1}, 53 (2006).

\bibitem{SFS} W. Guichard, M. Aprili, O. Bourgeois, T. Kontos,
  J. Lesueur, and P. Gandit, {Phase Sensitive Experiments in
    Ferromagnetic-Based Josephson Junctions}, Phys. Rev. Lett. {\bf
    90}, 167001 (2003).

\end{thebibliography}
\end{document}